\documentclass[12pt]{article}
\catcode`\@=11

\global\arraycolsep=2pt
\usepackage{amsbsy,amssymb,latexsym,amsfonts,amsmath}
\usepackage{graphicx,color,hyperref,braket}
\usepackage{tikz}
\usetikzlibrary{intersections, calc, arrows.meta}
\usepackage{graphicx}

\allowdisplaybreaks

\newcommand{\red}{\textcolor{red}}
\newcommand{\blue}{\textcolor{blue}}

\title{Infinitely many new renormalization group flows of minimal models from  non-invertible symmetries} 
\author{Yu Nakayama and Takahilo Tanaka}
\date{\today}

\begin{document}

\begin{flushright}
YITP-24-92

\end{flushright}

\vspace*{0.7cm}

\begin{center}
{ \Large  Infinitely many new renormalization group flows between Virasoro minimal models from  non-invertible symmetries}
\vspace*{1.5cm}\\
{Takahilo Tanaka and Yu Nakayama}

\end{center}

\vspace*{1.0cm}
\begin{center}

Yukawa Institute for Theoretical Physics,
Kyoto University, Kitashirakawa Oiwakecho, Sakyo-ku, Kyoto 606-8502, Japan

\vspace{1.5cm}
\end{center}

\begin{abstract}
Based on the study of non-invertible symmetries, we propose there exist infinitely many new renormalization group flows between Virasoro minimal models  $\mathcal{M}(kq + I, q) \to\mathcal{M}(kq-I, q)$ induced by $\phi_{(1,2k+1)}$. They vastly generalize the previously proposed ones $k=I=1$ by Zamolodchikov, $k=1, I>1$ by Ahn and L\"assig, $k=2, I=1$ by Martins, and $k=2$ with general $I$ by Dorey et al. All the other $\mathbb{Z}_2$ preserving renormalization group flows sporadically known in the literature (e.g. $\mathcal{M}(10,3) \to \mathcal{M}(8,3)$ studied by Klebanov et al) fall into our proposal (e.g. $k=3, I=1$). We claim our new flows give a complete understanding of the renormalization group flows between Virasoro minimal models that preserve a modular tensor category with the $SU(2)_{q-2}$ fusion ring.
\end{abstract}

\thispagestyle{empty} 

\setcounter{page}{0}

\newpage

\tableofcontents

\newpage

\section{Introduction}
While we believe we know everything about the Virasoro minimal models in two-dimensional conformal field theories, the renormalization group flow between them, in particular when one of them is non-unitary has been largely unknown. The Virasoro minimal model $\mathcal{M}(p,q)$ is characterized by two coprime integers $p$ and $q$, and we can compute all the conformal data for an arbitrary choice of $(p,q)$. It has been, however, a surprisingly difficult question to ask if they are connected by a renormalization group flow, given two randomly chosen integers $(p,q)$ and $(p',q')$, Our goal is to answer this question by using the non-invertible symmetries to classify the flows. Indeed, we will show that non-invertible symmetries give rise to infinitely many new renormalization group flows between two Virasoro minimal models.

Our infinitely many new flows
 $\mathcal{M}(kq + I, q) \to\mathcal{M}(kq-I, q)$ induced by $\phi_{(1, 2k + 1)}$ vastly generalize the previously proposed ones $k=I=1$ by Zamolodchikov \cite{Zamolodchikov:1987ti}, $k=1, I>1$ by Ahn \cite{Ahn:1992qi} and L\"assig \cite{Lassig:1991an}, $k=2, I=1$ by Martins\cite{Martins:1992ht,Martins:1992yk}, and $k=2$ with general $I$ by Dorey et al \cite{Dorey:2000zb}.\footnote{Martins \cite{Martins:1992ht,Martins:1992yk} and Dorey et al \cite{Dorey:2000zb} also studied the flows corresponding to $k=\frac{1}{2}$ with odd $q$. Since the preserved non-invertible symmetries become smaller, we study the generalization of half-integer $k$ separately in section \ref{halfk}.} It also encodes the other $\mathbb{Z}_2$ preserving renormalization group flow sporadically known in the literature (e.g. $\mathcal{M}(10,3) \to \mathcal{M}(8,3)$ studied by Klebanov et al \cite{Klebanov:2022syt}). With a slight twist, the renormalization group flows between multi-critical Lee-Yang fixed points $\mathcal{M}(2,q)$ studied in the literature \cite{Lencses:2022ira,Lencses:2023evr} fall into our proposal.  We claim that our flows give a complete understanding of the renormalization group flows between minimal models that preserve a modular tensor category (or more precisely modular fusion category) with the $SU(2)_{q-2}$ fusion ring.\footnote{The fusion ring $SU(2)_{q-2}$ is given by the fusion rule of the primary operators in the $SU(2)$ WZW model at level $q-2$. It is essentially angular momentum addition of $ J = 0, \frac{1}{2}, \cdots ,\frac{q-2}{2}$ with a $q$ dependent ``cap": $J_1+J_2 + J_3 \le q-2$. Given the fusion ring, the modular tensor category is further classified by the solutions of the Pentagon identity. The relation between the solutions of the Pentagon identity and the renormalization group invariants we will discuss can be found in \cite{Chang:2018iay}.}  

In our discussions, the non-invertible symmetries or categorical symmetries,\footnote{Recent applications of non-invertible symmetries in two-dimensional conformal field theories include \cite{Chang:2018iay,Lin:2019hks,Aasen:2020jwb,Thorngren:2021yso,Huang:2021nvb,Buican:2021uyp,Burbano:2021loy,Chang:2022hud,Lin:2022dhv,Lu:2022ver,Kaidi:2022cpf,Cheng:2022sgb,Chatterjee:2022jll,Lin:2023uvm,Jacobsen:2023isq,Choi:2023xjw,Haghighat:2023sax,Seiberg:2023cdc,Antinucci:2023ezl,Duan:2023ykn,Chen:2023jht,Nagoya:2023zky,Sinha:2023hum,Choi:2023vgk,Diatlyk:2023fwf,Cordova:2023qei,Grover:2023loq,Seiberg:2024gek,Copetti:2024rqj,Chatterjee:2024ych}. } which are realized by topological defect lines in two-dimensional conformal field theories, will play an analog of the 't Hooft anomaly. Similarly to the 't Hooft anomaly, their properties are preserved under the renormalization group flows. When the renormalization group invariants obtained from the non-invertible symmetries are different, they cannot be connected by the renormalization group flow that preserves the symmetries. We will show that our proposed renormalization group flows are not only consistent with the constraint but also give the actual flows between two minimal models with the same preserved non-invertible symmetries.

Some of our results may appeal to more physical intuitions. For instance, the preserved non-invertible symmetries in $\mathcal{M}(p,4) \to \mathcal{M}(p',4)$  flows are given by the $\mathbb{Z}_2$ Tambara-Yamagami modular tensor category in math terms or the duality defect line in physics terms. We know from mathematics that there are two distinct quantum dimensions consistent with the $\mathbb{Z}_2$ Tambara-Yamagami (or Ising) fusion rule. We also know from physics that the quantum dimensions are renormalization group invariants. Accordingly, there exist two intrinsically different duality defects in two-dimensional conformal field theories. Our new renormalization group flows know them and the flows are completely separated as long as the duality is preserved. One physical application of such a flow was to identify the fate of the non-supersymmetric Yukawa fixed point in two dimensions \cite{Nakayama:2022svf}. As a vast generalization, our new results should give us a new map to explore renormalization group flows in two-dimensional quantum field theories.

The organization of our paper is as follows. In section \ref{Virasoro}, we give a review of Virasoro minimal models and topological defect lines. In section \ref{claim}, we present our main claim of the new renormalization group flows and give supporting evidence from the study of the renormalization group invariants associated with the non-invertible symmetries. In section \ref{Examples}, we study several physically interesting examples of our new renormalization group flows. In section \ref{Conclusion}, we conclude with some discussions for future directions.


\section{Virasoro minimal models}\label{Virasoro}
Let us first state our conventions of Virasoro minimal models. We specify the Virasoro minimal model $\mathcal{M}(p, q)$ by two coprime integers $p$ and $q$. While our convention is more or less the same as the one in the yellow book \cite{DiFrancesco:1997nk}, one notable exception is they always assume $p > q$ in the yellow book, but we take $q$ to be a fixed integer, and we investigate the renormalization group flow that changes $p$. For instance, the well-known renormalization group flow from the tricritical Ising model to the critical Ising model is $\mathcal{M}(5,4) \to \mathcal{M}(3,4)$ in our paper rather than $\mathcal{M}(5,4) \to \mathcal{M}(4,3)$ as in the yellow book.

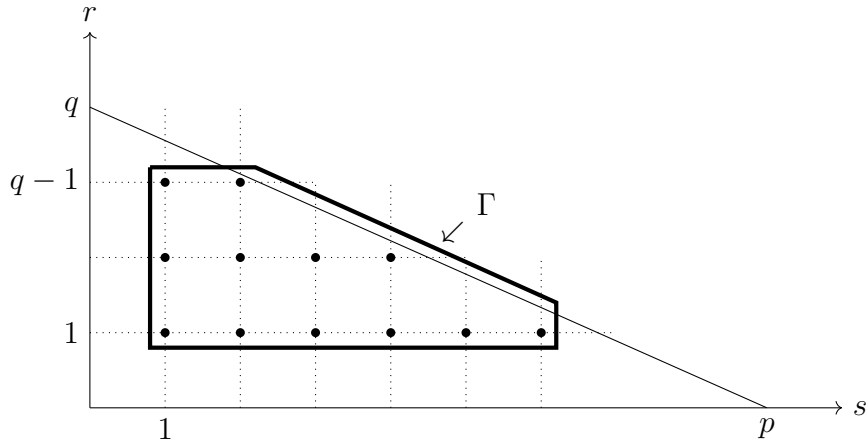
\begin{figure}[ht]
\centering
\begin{tikzpicture}
\draw[->](0,0)--(0, 5);
\draw[->](0, 0)--(10, 0);
\draw(0, 4)--(9, 0);
\fill(1, 3)circle(0.06);
\fill(2, 3)circle(0.06);
\fill(1, 2)circle(0.06);
\fill(2, 2)circle(0.06);
\fill(3, 2)circle(0.06);
\fill(4, 2)circle(0.06);
\fill(1, 1)circle(0.06);
\fill(2, 1)circle(0.06);
\fill(6, 1)circle(0.06);
\fill(5, 1)circle(0.06);
\fill(3, 1)circle(0.06);
\fill(4, 1)circle(0.06);
\draw(0, 5)node[above]{$r$};
\draw(10, 0)node[right]{$s$};
\draw(0, 4)node[left]{$q$};
\draw(9, 0)node[below]{$p$};
\draw[dotted](0, 3)--(3, 3);
\draw(0, 3)node[left]{$q - 1$};
\draw[dotted](0, 1)--(7, 1);
\draw(0, 1)node[left]{$1$};
\draw[dotted](1, 0)--(1, 4);
\draw(1, 0)node[below]{$1$};
\draw[dotted](6, 0)--(6, 2);
\draw[dotted](2, 0)--(2, 4);
\draw[dotted](3, 0)--(3, 3);
\draw[dotted](4, 0)--(4, 3);
\draw[dotted](5, 0)--(5, 2);
\draw[dotted](0, 2)--(5, 2);
\draw[ultra thick](0.8, 3.2)--(0.8, 0.8)--(6.2, 0.8)--(6.2, 1.4)--(2.2, 3.2)--(0.8, 3.2);
\draw(5, 2.5)node{\rotatebox{225}{$\to$} $\Gamma$};
\end{tikzpicture}
\caption{The Kac table and its fundamental domain $\Gamma$ of $\mathcal{M}(p, q)$. }
\label{zu1}
\end{figure}

The central charge of $\mathcal{M}(p, q)$ is 
\begin{align}
    c = \bar{c} = 1 - 6\dfrac{(p - q)^2}{pq}
\end{align}
and it has $\frac{(p - 1)(q - 1)}{2}$ (chiral) primary operators $\phi_{(r, s)}$ where $r$ and $s$ are the Kac indices whose ranges are $1 \le r < q, 1 \le s < p$. Two primary operators $\phi_{(r, s)}$ and $\phi_{(q -  r, p - s)}$ are identified: they have the same conformal weight 
\begin{align}
    h_{r, s} = h_{q - r, p - s} = \dfrac{(pr - qs)^2 - (p - q)^2}{4pq}. \label{dimension}
\end{align}
The fusion rule of the primary operators is given by
\begin{align}
    \phi_{(r, s)} \times \phi_{(m, n)} = \sum_{\substack{k = 1 + \lvert r - m \rvert\\ k + r + m = 1\mod2}}^{\min (r + m - 1, 2q - 1 - r - m)} \sum_{\substack{l = 1 + \lvert s - n \rvert\\ l + s + n = 1\mod2}}^{\min (s + n - 1, 2p - 1 - s - n)} \phi_{(k, l)}. \label{5}
\end{align}
For example, we can choose a fundamental domain $\Gamma$ of the Kac table, which specifies $\frac{(p - 1)(q - 1)}{2}$ distinct Kac indices as
\begin{align}
    \Gamma = \{ (r, s)\ |\ 1 \le r \le q - 1,\ 1 \le s \le p - 1,\ pr + qs < pq \}.
\end{align}
See Figure \ref{zu1}.

The Virasoro minimal models are further classified by the modular invariant partition functions. In this paper, we focus on those with A-series modular invariant partition functions (a.k.a. A-series minimal models). Since most of the renormalization group flows that we will discuss preserve the $\mathbb{Z}_2$ symmetry, which can be used to relate the A-series and D-series by orbifolding (when it can be gauged), most of the following discussions apply to the D-series minimal models. Exceptions are half-integer $k$ flow discussed in section \ref{halfk} because the deformation may not exist in D-series. {Furthermore, when we can gauge the $\mathbb{Z}_2 $ symmetry, we have fermionic minimal models \cite{Petkova:1988cy, Hsieh:2020uwb} as well. Our discussions also apply to their cases. Again, exceptions are half-integer $k$ flow discussed in section \ref{halfk}.} Our discussions do not directly apply to the E-series minimal models, however, and we need separate considerations. Some studies of the topological defect lines in E-series minimal models in view of the renormalization group flows can be found in \cite{Chang:2018iay,Nakayama:2022svf}.


\subsection{Non-invertible symmetries and topological defect lines}
In two-dimensional quantum field theories, non-invertible symmetries are synonyms of topological defect lines. In Virasoro minimal models with A-series modular invariant partition functions, it is widely believed that all the topological defect lines are given by the Verlinde lines.\footnote{To avoid trivial counterexamples, we here exclude space-time symmetries such as Poincar\'e transformation or space-time parity.} Verlinde lines $L_{(r,s)}$ have the same label $(r,s)$ as the (chiral) primary operators $\phi_{(r, s)}$ of the same theory, so  $\frac{(p-1)(q-1)}{2}$ of them are independent.

 The action of $L_{(r, s)}$ on a primary state $\ket{\phi_{(\rho, \sigma)}} = \phi_{(\rho, \sigma)}\ket{0}$ is given by
\begin{align}
    L_{(r, s)}\ket{\phi_{(\rho, \sigma)}} = \dfrac{S_{(r, s), (\rho, \sigma)}}{S_{(1, 1), (\rho, \sigma)}}\ket{\phi_{(\rho, \sigma)}}, \label{TDLonOP}
\end{align}
where $S_{(r, s), (\rho, \sigma)}$ is modular S-matrix.  The explicit expression of the S-matrix is
\begin{align}
    S_{(r, s), (\rho, \sigma)} = 2\sqrt{\dfrac{2}{pq}}(-1)^{1 + s\rho + r\sigma}\sin\Bigl(\pi\dfrac{p}{q}r\rho\Bigr)\sin\Bigl(\pi\dfrac{q}{p}s\sigma\Bigr),
\end{align}
which is real and symmetric.

As a special case, the action of $L_{(r, s)}$ on the vacuum $\ket{0} = \ket{\phi_{(1, 1)}}$ is given by
\begin{align}
    L_{(r, s)}\ket{0} = d_{(r, s)}\ket{0} = \dfrac{S_{(r, s), (1, 1)}}{S_{(1, 1), (1, 1)}}\ket{0},  \label{quantumdim}
\end{align}
where the eigenvalue $d_{(r, s)}$ is called quantum dimension of $L_{(r, s)}$. The salient property of the quantum dimension is that it is a renormalization group invariant when $L_{(r,s)}$ commutes with the deforming operators as we will see.

The Verlinde lines satisfy the same fusion rule as (chiral) primary operators:
\begin{align}
    L_{(r, s)} \times L_{(m, n)} &=  \sum N^{(k,l)}_{(r,s)(m,n)} L_{(k,l)} \cr
    &= \sum_{\substack{k = 1 + \lvert r - m \rvert\\ k + r + m = 1\mod2}}^{\min (r + m - 1, 2q - 1 - r - m)} \sum_{\substack{l = 1 + \lvert s - n \rvert\\ l + s + n = 1\mod2}}^{\min (s + n - 1, 2p - 1 - s - n)} L_{(k, l)}. \label{fusionL} 
\end{align}
The fusion coefficients $ N^{(k,l)}_{(r,s)(m,n)}$ are related to the modular S-matrix by the Verlinde formula:
\begin{align}
N_{ab}^c = \sum_d \frac{S_{ad} S_{bd} S_{dc} }{S_{0d}}, \label{Verlinde}
\end{align}
where we collectively denote $(r,s)$ by $a$ etc. This short-hand collective notation is sometimes used hereafter.

\subsection{Invertible \texorpdfstring{$\mathbb{Z}_2$}{TEZT} symmetry}
All the Virasoro minimal models with A-series modular invariance (except when $p$ or $q$ is $2$) possess a special topological defect line $L_{(q - 1, 1)}$ corresponding to the invertible $\mathbb{Z}_2$ symmetry. It is invertible because the fusion rule (see e.g. \eqref{fusionL}) gives $L_{(q - 1, 1)} \times L_{(q - 1, 1)} = 1$. 

 Under the action of $L_{(q - 1, 1)}$, $\mathbb{Z}_2$ charge of the primary operators $\phi_{(r, s)}$ can be summarized as  
\begin{align}
\begin{split}
    r-1 \ &\mod 2 \ \ (\text{when $q$ is even}) \\
    s-1 \ &\mod 2  \ \ (\text{when $p$ is even}) \\
    r+s-2 \ &\mod 2  \ \ (\text{when $q$ and $p$ are both odd})  .
\end{split}
\end{align}
See e.g. \cite{Lassig:1991an}.
The $\mathbb{Z}_2$ symmetry is compatible with the fusion rule of the primary operators and operator identifications. If we gauge the $\mathbb{Z}_2$ symmetry (or orbifold it in the conformal field theory language), A-series minimal models are exchanged with the D-series minimal models unless it is anomalous.

At this point, we note that the $\mathbb{Z}_2$ symmetry is anomalous when $p$ and $q$ are both odd. One way to see this is to compute the quantum dimensions of $L_{(q - 1, 1)}$, which is $-1$ rather than $+1$.  As we will review in the next subsection, the quantum dimension is a renormalization group invariant and it gives a selection rule of the renormalization group flow as the 't Hooft anomaly. We can also compute the spin contents of the defect Hilbert space, which will be again reviewed in the next subsection, and see that it is $\pm\frac{1}{4}$ rather than $0,\pm \frac{1}{2}$. As discussed in \cite{Chang:2018iay} this is a clear signal of the 't Hooft anomaly. Let us also recall that there is no D-series modular invariant partition function when $p$ and $q$ are both odd. This means that we cannot gauge the $\mathbb{Z}_2$ invertible symmetry and it is consistent with the existence of the anomaly. 

In this paper, we focus on the renormalization group flow that preserves the $\mathbb{Z}_2$ invertible symmetry. To be more precise, our renormalization group flow will preserve a larger modular tensor category with the $SU(2)_{q-2}$ fusion ring. This modular tensor category includes the $\mathbb{Z}_2$ invertible symmetry but gives extra $q-3$ non-invertible symmetries.  In the next subsection, we will review how the preserved non-invertible symmetries will constrain the renormalization group flows.

\subsection{Renormalization group invariants}\label{RGinv}
The salient feature of the topological defect lines regarded as non-invertible symmetries is that they give renormalization group invariants when they are preserved along the flow. Let us first discuss the condition that the topological defect lines are preserved along the renormalization group flow in the general setup.

Suppose we are at the ultraviolet conformal field theory with a topological defect line $L_a$. We deform the theory by a relevant operator $\phi_b$. {More generally we have to consider a collection of relevant operators $\{ \phi_b \}$ to reach the desired infrared fixed point. In such a case, we assume that the following conditions are met with all the operators we need.} Now, we say that $L_a$ is preserved along the renormalization group flow when they commute with $\phi_b$:  $L_{a} \phi_b = \phi_b L_{a}$ on the cylinder. More explicitly, they must satisfy
\begin{align}
L_a \phi_b \ket{\Phi} = \phi_b L_a \ket{\Phi} 
\end{align}
on any state $\ket{\Phi}$ on the cylinder. In unitary conformal field theories, it is sufficient to check it on the vacuum $\ket{0} = \ket{\phi_{(1, 1)}}$ \cite{Chang:2018iay}, but in non-unitary conformal field theories, we may have to study all the states. We will, however, explicitly see that in our discussions we do not gain any new constraint by studying all the states rather than the vacuum alone.

In the mathematics language, topological defect lines give a realization of the modular tensor category. As a modular tensor category, the physical properties of the topological defect lines are severely constrained. It is believed that the finite-rank modular tensor categories have no continuous parameters (so-called rigidity). Therefore, when the topological defect lines are preserved, we expect that the properties of the topological defect lines as a modular tensor category should be preserved by regarding the renormalization group flow as a continuous deformation.

Some properties of the topological defect lines are directly related to the 't Hooft anomaly i.e. an obstruction to gauge the symmetry, in which case it is not necessary to assume the non-existence of the continuous deformation of the modular tensor category to deduce the renormalization group invariance of the modular tensor category. It is simply the 't Hooft anomaly matching. 

In this paper, we focus on two such properties. One is the quantum dimensions of topological defect lines and the other is the spin contents of the defect Hilbert space. See e.g. \cite{Kikuchi:2021qxz,Kikuchi:2022jbl,Kikuchi:2022gfi,Kikuchi:2022biw} for studies of these properties in various examples.

We have already introduced the quantum dimensions of the topological defect lines $L_a$ in \eqref{quantumdim}. 
 The renormalization group invariance of the quantum dimensions can be understood as follows. Recall that the fusion of the topological defect lines is
\begin{align}
    L_a \times L_b = \sum_{c} N_{ab}^{c}L_c,
\end{align}
where $N_{ab}^{c}$ is the fusion coefficients. 

Let us now take the vacuum expectation value of this equality: $\langle L_a \rangle := \langle 0|L_a | 0 \rangle$, where $|0\rangle = |\phi_{(1,1)} \rangle$ on the cylinder. The topological defect lines are topological, and we can place them far away. Then we may assume that the expectation value is factorized\footnote{Potentially, this argument can fail because we are applying the idea developed in unitary quantum field theories to non-unitary conformal field theories. It, however, turns out that the quantum dimensions that we can compute explicitly in non-unitary minimal models do satisfy the constraints obtained from this cluster decomposition ansatz.}
\begin{align}
    \langle L_a \rangle \times \langle L_b \rangle = \sum_{c} N_{ab}^{c} \langle L_c \rangle.
\end{align}
The solutions of these quadratic equations turn out to be discrete, so it is invariant under the continuous renormalization group flows.

As the simplest example, let us consider the $\mathbb{Z}_2$ invertible symmetry with the fusion rule $L_{(q - 1, 1)} \times L_{(q - 1, 1)} = L_{(1, 1)} (= 1)  $. 
 By taking the vacuum expectation value, we can immediately see that $\langle L_{(q - 1, 1)} \rangle = \pm 1$. The sign choice corresponds to the anomaly of the $\mathbb{Z}_2$ invertible symmetry.\footnote{Whether one can choose the sign depends on the constraint from the other fusion rules. In Appendix C of \cite{Nakayama:2022svf}, we can find the discussion of why the minus sign is still consistent with the $\mathbb{Z}_2$ selection rule in correlation functions.} In this case, the renormalization group invariance of the quantum dimension is equivalent to the 't Hooft anomaly matching.

The other renormalization group invariants we will discuss are the spin contents of the defect Hilbert space associated with the preserved topological defect line $L_{a}$. To introduce the notion of the defect Hilbert space and the spin contents, we begin with the torus partition function with the insertion of $L_{a}$ in the spatial direction.
\begin{align}
Z_{L_a}(\tau, \bar{\tau}) &= \mathrm{Tr} \left( L_{a} q^{L_0-c/24}\bar{q}^{\bar{L}_0-c/24} \right) \cr
&= \sum_b \frac{S_{a b}}{S_{0b}} \chi_b(\tau) \bar{\chi}_b(\bar{\tau}) ,
\end{align}
where $q= e^{2\pi i \tau}$ and $\chi_b$ is the Virasoro character of the primary operator $\phi_{b}$.

Let us perform the S-modular transformation ($\tilde{\tau} = -\tau^{-1}$) to regard the partition function as the sum over the defect Hilbert space $\mathcal{H}_{L_{a}}$:
\begin{align}
Z_{L_a}(\tau, \bar{\tau}) &= \sum_{b, c, d} \frac{S_{a b}}{S_{0b}} S_{bc} S_{bd} \chi_c(\tilde{\tau}) \bar{\chi}_d(\bar{\tilde{\tau}}) \cr
& = \sum_{c, d} N_{cd}^{a} \chi_c(\tilde{\tau}) \bar{\chi}_d(\bar{\tilde{\tau}}) \cr
&= \mathrm{Tr}_{\mathcal{H}_{L_{a}}} \left(  \tilde{q}^{L_0-c/24}\bar{\tilde{q}}^{\bar{L}_0-c/24} \right),
\end{align}
where we have used the Verlinde formula \eqref{Verlinde}. In the last line, the topological defect line is inserted in the time direction, and the quantization condition on the spatial circle is modified.
The spin contents of the defect Hilbert space $\mathcal{H}_{L_{a}}$ are defined by the set of ``Lorents spin" $\{ h_c - h_d \ (\mathrm{mod} \ \mathbb{Z}) \}$ evaluated over non-zero $N_{cd}^{a}$.

Let us now argue that the spin contents of preserved topological defect lines are renormalization group invariants (see e.g. \cite{Chang:2018iay}). The idea is that the topological defect lines are topological and the relevant deformation preserves the $U(1)$ rotational symmetry of the quantum field theory. It implies that the twisted quantization condition on the defect Hilbert space cannot change in a discrete manner. More precisely, in \cite{Kikuchi:2022jbl,Kikuchi:2022rco} it is claimed that the spin content of the infrared theory must be a subset of the ultraviolet theory. The spin contents of the infrared theory can be a subset because the states might become heavy and decouple.

\section{New renormalization group flows}\label{claim}
In this section, based on the study of non-invertible symmetries, we propose infinitely many new renormalization group flows in Virasoro minimal models. We give some formal checks of the agreement of renormalization group invariants under the flow. Examples will be given in the next section.

\subsection{Renormalization group flow \texorpdfstring{ $\mathcal{M}(kq + I, q) \to\mathcal{M}(kq-I, q)$}{TEXT}}

The main claim of our paper is there exist infinitely many renormalization group flows $\mathcal{M}(kq + I, q) \to\mathcal{M}(kq-I, q)$ induced by the $\phi_{(1, 2k + 1)}$ deformation, which satisfy the constraint from the non-invertible symmetries. We further claim that the renormalization group flow is complete when we preserve a modular tensor category with the  $SU(2)_{q-2}$ fusion rule that includes the $\mathbb{Z}_2$ invertible symmetry. By complete, we mean that all the possible flows that preserve a modular tensor category with the $SU(2)_{q-2}$ fusion ring can be represented by a repeated use of our proposed flows $\mathcal{M}(kq + I, q) \to\mathcal{M}(kq-I, q)$.
For concreteness, we assume $k$ and $I$ are both integers in this subsection. The generalization for non-integer $k$ (and $I$) will be discussed in the next subsection. The half-integer case does not preserve the modular tensor category with the $SU(2)_{q-2}$ fusion ring.

We first show that the $\phi_{(1, 2k + 1)}$ deformation preserves the $SU(2)_{q-2}$ subcategory $L_{(r, 1)}$ $(i= 1, \cdots, q-1)$ out of the non-invertible symmetries of the undeformed theory $\mathcal{M}(kq + I, q)$. For this purpose, as reviewed in section \ref{RGinv}, we should check $L_{(r,1)}$ and $\phi_{(1,2k+1)}$ commute on the cylinder.

Let us start with the fusion rule \eqref{5}. We have
\begin{align} 
    \phi_{(1, 2k + 1)} \times \phi_{(\rho, \sigma)} = \sum_{\substack{l = 1 + \lvert 2k + 1 - \sigma \rvert\\ l + \sigma = 0\mod2}}^{\min (2k + \sigma, 2(kq + I) - 2- 2k - \sigma)} \phi_{(\rho, l)} = \sum_{l \in \Lambda} \phi_{(\rho, l)}
\end{align}
in $\mathcal{M}(kq + I, q)$, where we have defined $\Lambda$ as 
\begin{align}
    \Lambda = \{\ l\ |\ 1 \le l \le \min (2k + \sigma, 2(kq + I) - 2- 2k - \sigma),\ l + \sigma = 0 \ \ \mathrm{mod} \ 2 \ \},
\end{align}
and then we can compare the action of $L_{(r, 1)}\phi_{(1, 2k + 1)}$ and $\phi_{(1, 2k + 1)} L_{(r, 1)}$  on arbitrary primary states $\ket{\phi_{(\rho, \sigma)}}$:
\begin{align}
    L_{(r, 1)}\phi_{(1, 2k + 1)}\ket{\phi_{(\rho, \sigma)}} &= L_{(r, 1)}\phi_{(1, 2k + 1)}\phi_{(\rho, \sigma)}\ket{0} = \sum_{l \in \Lambda} L_{(r, 1)}\phi_{(\rho, l)}\ket{0}\notag\\
    &= \sum_{l \in \Lambda} L_{(r, 1)}\ket{\phi_{(\rho, l)}} = \sum_{l \in \Lambda} \dfrac{S_{(r, 1), (\rho, l)}}{S_{(1, 1), (\rho, l)}}\ket{\phi_{(\rho, l)}}\\
    \phi_{(1, 2k + 1)}L_{(r, 1)}\ket{\phi_{(\rho, \sigma)}} &= \dfrac{S_{(r, 1), (\rho, \sigma)}}{S_{(1, 1), (\rho, \sigma)}}\phi_{(1, 2k + 1)}\ket{\phi_{(\rho, \sigma)}} = \dfrac{S_{(r, 1), (\rho, \sigma)}}{S_{(1, 1), (\rho, \sigma)}}\phi_{(1, 2k + 1)}\phi_{(\rho, \sigma)}\ket{0}\notag\\
    &= \sum_{l \in \Lambda} \dfrac{S_{(r, 1), (\rho, \sigma)}}{S_{(1, 1), (\rho, \sigma)}}\phi_{(\rho, l)}\ket{0} = \sum_{l \in \Lambda} \dfrac{S_{(r, 1), (\rho, \sigma)}}{S_{(1, 1), (\rho, \sigma)}}\ket{\phi_{(\rho, l)}}.
\end{align}
Because $l + \sigma$ is even in the above sum,
\begin{align}
    \left.\dfrac{S_{(r, 1), (\rho, l)}}{S_{(1, 1), (\rho, l)}} \middle/ \right.\dfrac{S_{(r, 1), (\rho, \sigma)}}{S_{(1, 1), (\rho, \sigma)}} &= \left.(-1)^{(r - 1)l}\dfrac{\sin(\pi\frac{kq + I}{q}r\rho)}{\sin(\pi\frac{kq + I}{q}\rho)} \middle/ \right.(-1)^{(r - 1)\sigma}\dfrac{\sin(\pi\frac{kq + I}{q}r\rho)}{\sin(\pi\frac{kq + I}{q}\rho)}\notag\\
    &= 1 \label{sum}
\end{align} 
and hence $L_{(1, 1)}, L_{(2, 1)}, \cdots, L_{(q - 2, 1)}, L_{(q - 1, 1)}$ commute with $\phi_{(1,2k+1)}$. This means that $SU(2)_{q-2}$ subcategory generated by $L_{(r,1)}$ is preserved under the renormalization group flow induced by the $\phi_{(1, 2k + 1)}$ deformation. In particular, we stress that $L_{(q - 1, 1)}$ generates the invertible $\mathbb{Z}_2$ symmetry of the minimal models, so the renormalization group flow is $\mathbb{Z}_2$ symmetric. 

In the above discussions, we realize that $\phi_{(1,2k+1)}$ can be replaced with $\phi_{(1,2l+1)}$ for any integer $l$ such that  $\phi_{(1,2l+1)}$ is in the Kac table. The physical significance of choosing $l=k$ is that then $\phi_{(1,2k+1)}$ will be the least relevant deformation within $\{\phi_{(1,2l+1)} \}$ if $I<k$.  When $k>1$, we typically have to introduce the other more relevant deformation $\phi_{(1,2l+1)}$ (with $l<k$) and fine-tune the deformation parameters to get to the critical point.\footnote{Generally we have to fine-tune $k-1$ parameters.} When $I>k$, we need fewer fine-tunings to reach the fixed point dictated by the proposed flow.

Let us now show that the quantum dimensions of all the preserved topological defect lines match under the proposed renormalization group flows $\mathcal{M}(kq + I, q) \to\mathcal{M}(kq-I, q)$.\footnote{The matching of the quantum dimensions when $k=1$ and $k=2$, corresponding to the previously known renormalization group flows, was discussed in \cite{Kikuchi:2021qxz,Kikuchi:2022biw}.} The proof follows by the direct calculation.
The quantum dimension $d_{(r, 1)}$ of $L_{(r, 1)}$ in $\mathcal{M}(kq + I, q)$ is $(- 1)^{r + 1}\sin(\pi\frac{kq + I}{q}r)/\sin(\pi\frac{kq + I}{q})$ and that in $\mathcal{M}(kq - I, q)$ is $(- 1)^{r + 1}\sin(\pi\frac{kq - I}{q}r)/\sin(\pi\frac{kq - I}{q})$. Its ratio is 
\begin{align}
    \dfrac{\sin(\pi\frac{kq + I}{q}r)}{\sin(\pi\frac{kq - I}{q}r)}\dfrac{\sin(\pi\frac{kq - I}{q})}{\sin(\pi\frac{kq + I}{q})} = \dfrac{\sin(\pi\frac{I}{q}r)}{\sin(- \pi\frac{I}{q}r)}\dfrac{\sin(- \pi\frac{I}{q})}{\sin(\pi\frac{I}{q})} = 1. \label{qdratio}
\end{align}

We can further argue that the constraint from the quantum dimensions is sufficiently non-trivial. We observe that because $kq + (\pm I + q) = (k + 1)q \pm I$, $I$ takes a value in $1 \le I \le q - 1$, and $kq \pm I$ and $q$ are coprime, our proposed flows can be classified by $\varphi(q)$ kinds of $I$, where $\varphi(q)$ is Euler function. {Now we want to see that each connected flows have different quantum dimensions.} For this purpose, let us examine the quantum dimensions of $L_{(2,1)}$ in $\mathcal{M}(p,q)$ in particular; the constraint from the other preserved topological defect lines is typically not as strong. The explicit evaluation of the formula \eqref{quantumdim} gives $d_{(2,1)} = -2 \cos\left(\frac{p}{q}\pi\right)$ and we immediately observe $d_{(2,1)}$ takes a distinct value for different $p$ ($\mathrm{mod} \ q$), given $q$. Thus, the quantum dimensions do distinguish our renormalization group flows.

From the general argument in section \ref{RGinv}, under the new renormalization group flow, the spin contents of the preserved line  $L_{(r,1)}$ $(r = 1, \cdots, q - 1)$ should remain the same or become a subset.
While it may be non-trivial to check the agreement of all the spin contents in the most generic cases, one can systematically check the agreement in the following way. 

We first realize that $N_{(s, m)(t, n)}^{(r, 1)} \chi_{(s, m)} \bar{\chi}_{(t, n)}$ is non-zero only when $m = n$. Furthermore, the fusion coefficients necessary for us do not depend on $m$, so we can first determine the non-zero entry of $N_{(s, m)(t, n)}^{(r, 1)}= \delta_{mn} N_{(s, 1)(t, 1)}^{(r, 1)}$ from $SU(2)_{q - 2}$ fusion rule, and then we can compute the spin of the defect Hilbert space from \eqref{dimension}. 
In this way, it is possible to systematically compute the spin contents of $L_{(r,1)}$ and see the agreement under the proposed renormalization group flow. We will show some examples in section \ref{Examples}.



Now let us show that the spin contents of $\mathcal{H}_{L_{(2,1)}}$ in  $\mathcal{M}(p,q)$ are preserved 
 under the proposed renormalization group flows. We first observe that the only non-zero fusion coefficients are $N_{(t, n) (t + 1, m)}^{(2, 1)}$ and $N_{(t + 1, m) (t, m)}^{(2, 1)}$. By computing $\pm (h_{(t + 1, m)} - h_{(t, m)})$, we find the spin contents of $\mathcal{H}_{L_{(2,1)}}$ are $\pm\frac{p}{q} \frac{1 + 2t}{4}$ (mod $\frac{1}{2} \mathbb{Z}$) with $t = 1, \cdots, q - 2$. Thus, the spin contents of $\mathcal{H}_{L_{(2,1)}}$ in $\mathcal{M}(kq + I, q)$ contain those in $\mathcal{M}(kq - I, q)$.

{We further want to see that the spin contents of $\mathcal{H}_{L_{(2,1)}}$ distinguish the renormalization group flows as the quantum dimensions. 
We observe that $\frac{p}{q} \frac{1+2t}{4}$ never equals to $\frac{p}{q} \frac{1}{4}$ mod $\frac{\mathbb{Z}}{2}$, so $\frac{p}{q} \frac{1}{4}$  is always missing in the spin content. (Indeed, if this were true, we would have $\frac{p}{q}l = \mathbb{Z}$, but since $p$ and $q$ are coprime, it should be a contradiction.) On the other hand, the missing ones (except $\frac{1}{4q}$ and $\frac{1}{2q}$) always appear in $p=2q-1$ or $p=2q-2$ as long as $q>3$, so the spin contents of $p=2q-1$ and $p=2q-2$ are different from the other $p$'s (mode $q$). We can repeat the analysis with the other $p=2q-s$ with $s=3, 4, \cdots, q-1$ to realize that they all have different spin contents. The exceptional but easier $q = 3$ case can be treated separately (see section \ref{M3}).

The renormalization group invariants cannot tell which of $\mathcal{M}(kq+I,q)$ or  $\mathcal{M}(kq-I,q)$ is the ultraviolet theory. To address this question, let us mention two evidence that  $\mathcal{M}(kq+I,q)$ is the ultraviolet theory. The first evidence is the $c_{\mathrm{eff}}$ theorem \cite{Castro-Alvaredo:2017udm}. While $\mathcal{M}(p,q)$ can be non-unitary, it is believed that as long as the deformation preserves the $\mathcal{PT}$-symmetry (see e.g. \cite{Bender:2023cem} for a review), the renormalization group flow shows the monotonicity with respect to the effective central charge $c_{\mathrm{eff}} = 1 -\frac{6}{pq}$ along the flow. If we accept the $c_{\mathrm{eff}}$ theorem (with the assumption our flows preserve the $\mathcal{PT}$-symmetry), we may conclude $\mathcal{M}(kq+I,q)$ is the ultraviolet theory.\footnote{One caveat here is that we have not verified our renormalization group flow preserves the $\mathcal{PT}$ symmetry. It is an interesting question to see if the $\mathcal{PT}$ symmetry can be realized as a topological defect line.} 

The other evidence comes from the number of (singlet) relevant deformations \cite{Gukov:2015qea}. It was argued there that along the renormalization group flow, the number of (singlet) relevant deformations should decrease.
Intuitively, if we deform the theory by relevant perturbation, the deformed theory becomes less critical and we expect less fine-tuning is needed to reach the criticality. In \cite{Gukov:2016tnp}, it was put on firmer ground by studying the topology of the renormalization group flow. Again once we accept this conjecture, we realize $\mathcal{M}(kq+I,q)$ is the ultraviolet theory because it has a larger number of relevant singlet operators. Note also that if our constraint from the non-invertible symmetries did not exist, we could have found counterexamples of the conjecture. For instance in the would-be flow $\mathcal{M}(7,4) \to \mathcal{M}(5,4)$, the number of relevant singlet operators did not decrease. We have, however, forbidden the flow by matching the non-invertible symmetries. In this discussion, it is important to note that by singlet, we mean the operators that commute with all the non-invertible symmetries $\{L_{(r,1)} \}$ whose fusion ring is $SU(2)_{q-2}$ than just those that commute with the $\mathbb{Z}_2$ invertible symmetry.


Here are some historical comments. In a classic paper \cite{Zamolodchikov:1987ti}, Zamolodchikov studied the renormalization group flow $k=I=1$, which turns out to be integrable \cite{Zamolodchikov:1987jf}. This case is most physically interesting because it describes the flow between unitary minimal models. Subsequently, Ahn \cite{Ahn:1992qi} and L\"assig \cite{Lassig:1991an} proposed the generalization to the non-unitary minimal models independently, corresponding to $k=1, I>1$. Later, Dorey, Dunning and Tateo \cite{Dorey:2000zb} (see also \cite{Martins:1992ht,Martins:1992yk,Ravanini:1994pt}) studied the case with $k=2$ from the viewpoint of non-linear integral equations. Our approach, based on the non-invertible symmetries, also provides us with new insights into the previously known flows.

\subsection{Half-integer \texorpdfstring{$k$}{TEXT}}\label{halfk}
Our proposed renormalization group flow
 $\mathcal{M}(kq + I, q) \to \mathcal{M}(kq - I, q)$ induced by $\phi_{(1, 2k + 1)}$
 formally makes sense even if $k$ is a half-integer, where $I$ takes a half-integer or integer depending on $q$. The main difference is that $2k+1$ is even so not all the topological defect lines $L_{(r, 1)}$ commute with $\phi_{(1, 2k + 1)}$, but only the subcategory spanned by $L_{(2\ell + 1, 1)}$, where $\ell$ is an integer, is preserved. This restriction can be explicitly seen in \eqref{sum}, where for half-integer $k$, $l+\sigma$ is odd (rather than even for integer $k$ case discussed there), so we have to assume $r$ is an odd integer. This subcategory has the fusion ring of $SO(3)_{[\frac{q}{2}]-1}$.
 
If we restrict ourselves to these topological defect lines, all the discussions above can be repeated. In particular, the consistency of the proposed renormalization group flow with the preserved non-invertible symmetry holds. For instance, the invariance of the  quantum dimensions of $L_{(r,1)}$ can be checked as in \eqref{qdratio}, but with half-integer $k$, the ratio becomes $(-1)^{r+1}$. Thus only the preserved lines with $r=2\ell+1$ have the same quantum dimensions as claimed. The strongest constraint from the spin contents comes from $L_{(3,1)}$ (rather than $L_{(2,1)}$ which does not exist). In $\mathcal{M}(p,q)$, the spin contents of $\mathcal{H}_{L_{(3,1)}}$ are $\{0,\pm \frac{p}{q}(\ell+1) \}$ ($\ell=1, \cdots, q-3$), and we can check it is consistent with our flow $\mathcal{M}(kq + I, q) \to \mathcal{M}(kq - I, q)$ with half-integer $k$.

Some specific renormalization group flows have been studied in the literature.
{The particular case with $k=\frac{1}{2}$ was discussed in Dorey et al \cite{Dorey:2000zb}, but note $(p,q)$ there was swapped compared with our unified notation so that their $\phi_{(1,2)}$ is our $\phi_{(2,1)}$. The non-invertible symmetries, in this previously known case, were studied in \cite{Kikuchi:2022biw}. Again, we generalize their works by letting $k$ be an arbitrary half-integer.}

Let us comment on the fate of the $\mathbb{Z}_2$  invertible symmetry under the half-integer $k$ flow.
 When $k$ is a half-integer and $q$ is odd, $\phi_{(1, 2k + 1)}$ is odd under the $\mathbb{Z}_2$ invertible symmetry, so the preserved modular tensor category does not include $L_{(q - 1, 1)}$. Nevertheless, after the flow, we do recover the entire modular tensor category with the $SU(2)_{q-2}$ fusion ring, including the invertible $\mathbb{Z}_2$ symmetry. They must be emergent symmetries. While this must be the case since this half-integer $k$ flow allows us to change the (non-)anomalous nature of the $\mathbb{Z}_2$ symmetry, the microscopic details of the renormalization group flow may be of interest and needs further study.

\section{Some examples}\label{Examples}

In this section, we study explicit examples of $\mathcal{M}(kq+I,q) \to \mathcal{M}(kq-I,q)$ flows for $q=3,4,5$, emphasizing physical interpretations. As a potentially exceptional case, we then examine $q=2$ separately.


\subsection{\texorpdfstring{$\mathcal{M}(3k+I,3) \to \mathcal{M}(3k-I,3)$}{TEXT} and anomaly matching}\label{M3}

In this case, our proposal with integer $k$ boils down to the conclusion that there exists distinct renormalization group flows among  $\mathcal{M}(2l+1,3)$ and among $\mathcal{M}(2l,3)$. The only non-trivial topological defect line preserved by the $\phi_{(1, 2k + 1)}$ deformation is $L_{(2,1)}$ that generates the invertible $\mathbb{Z}_2$ symmetry. The constraint from the quantum dimensions as well as spin contents will be equivalent to the 't Hooft anomaly matching. We recall that the $\mathbb{Z}_2$ invertible symmetry is anomalous in  $\mathcal{M}(2l+1,3)$ and non-anomalous in $\mathcal{M}(2l,3)$ so they cannot mix under the $\mathbb{Z}_2$ preserving renormalization group flow.

{While the quantum dimensions of $L_{(2,1)}$ can be directly computable from \eqref{quantumdim}, the spin contents of $\mathcal{H}_{L_{(2,1)}}$ can be computed by the algorithm mentioned in section \ref{RGinv}.}
As argued there, non-zero entry of $N_{(s, m)(t, n)}^{(2, 1)}$ implies $m = n$ and $(s, t) = (2, 1)\ \mathrm{or}\ (1, 2)$, which have the opposite spins. From the dimensions of primary operators \eqref{dimension}, we find the spin of the defect Hilbert space $\mathcal{H}_{L_{(2,1)}}$ is given by  $\pm(h_{(2, m)}-h_{(1, m)}) = \pm\frac{1}{4}(p - 2m)$ with $ 1\le m \le p - 1$. The spin contents are therefore $ \{ \pm \frac{1}{4} \}$ when $p$ is odd and $\{0, \pm \frac{1}{2} \} $ 
 when $p$ is even. 

We have summarized the quantum dimensions and spin contents of $\mathcal{M}(p,3)$ models in Table \ref{tableM3} and Table \ref{tablespin3}. Note that for the anomalous $\mathbb{Z}_2$ invertible symmetry, the quantum dimension is $-1$ and the spin content is $\{ \pm \frac{1}{4} \}$, which is consistent with the general argument \cite{Chang:2018iay}.

\begin{table}[ht]
\begin{center}
\begin{tabular}{|c|c|c|c|c|c|c|c|c|c|}\hline
     $p$&4&5&7&8&10&11&13&14 & $\cdots$ \\\hline
     $d_{(2, 1)}$ & $1$ &  $-1$ & $-1$ & $1$ & $1$ & $-1$ & $-1$ &$1$ &$ \cdots$  \\\hline
\end{tabular}
\caption{The quantum dimensions of $L_{(2, 1)}$ in $\mathcal{M}(p , 3)$ series.}\label{tableM3}
\end{center}
\end{table}

\begin{table}[ht]
\begin{center}
\begin{tabular}{|c|c|c|c|c|c|c|c|c|c|}\hline
     $p$&4&5&7&8&10&11&13&14 & $\cdots$ \\\hline
     $\mathcal{H}_{L_{(2, 1)}}$ & $0, \pm \frac{1}{2}$ &  $ \pm \frac{1}{4}$ & $ \pm \frac{1}{4}$ & $0, \pm \frac{1}{2}$ & $0, \pm \frac{1}{2}$ & $ \pm \frac{1}{4}$ & $ \pm \frac{1}{4}$ &$0, \pm \frac{1}{2}$ &$ \cdots$  \\\hline
\end{tabular}
\caption{The spin contents of $\mathcal{H}_{L_{(2, 1)}}$ in $\mathcal{M}(p , 3)$ series.}\label{tablespin3}
\end{center}
\end{table}

We would like to note that in the literature \cite{Fei:2014xta,Klebanov:2022syt}, one of the flows in this class $\mathcal{M}(10,3) \to \mathcal{M}(8,3)$ induced by $\phi_{(1,7)}$ ($k=3, I=1 $) was used to study the Landau-Ginzburg-like Lagrangian constructions of non-unitary minimal models. The further flow $\mathcal{M}(8,3) \to \mathcal{M}(4,3)$ induced by $\phi_{(1,5)}$ ($k=2, I=2 $) can be straightforwardly realized in their Landau-Ginzburg-like Lagrangian. Another flow $\mathcal{M}(7,3) \to \mathcal{M}(5,3)$ induced by $\phi_{(1, 5)}$ ($k = 2, I=1 $) was studied 
in \cite{Delouche:2023wsl} from the truncated conformal space approach.

Let us briefly discuss the case with half-integer $k$. With the half-integer $k$, $\phi_{(1,2k+1)}$ deformation does not preserve any non-trivial topological defect lines in $\mathcal{M}(p,3)$, so the constraint is none. We, however, observe that some of the predicted flow such as $\mathcal{M}(10,3) \overset{\phi_{(1, 6)}}{\longrightarrow} \mathcal{M}(5,3) \overset{\phi_{(1, 4)}}{\longrightarrow} \mathcal{M} (4,3)$ may give an interesting Lagrangian description of $\mathcal{M}(5,3)$, given the Lagrangian description of $\mathcal{M}(10,3)$ is available. Indeed, the flow has some theoretical interest because of the (dis)appearance of the anomalous $\mathbb{Z}_2$ symmetry of $\mathcal{M}(5,3)$ 
 in the middle of the renormalization group flow.

\subsection{\texorpdfstring{$\mathcal{M}(4k+I,4) \to \mathcal{M}(4k-I,4)$}{TEXT} and duality defects}

Here, under the $\phi_{(1, 2k + 1)}$ deformations with integer $k$, the preserved topological defect lines are $L_{(1,1)}, L_{(2,1)}$ (sometimes called $\mathcal{N}$) and $L_{(3,1)}$ (sometimes called $\eta$) and generate the $\mathbb{Z}_2$ Tambara-Yamagami modular tensor category. 
There are $\varphi(4) = 2$ distinct renormalization group flows predicted from our proposal.

Let us first study the quantum dimensions of $L_{(2, 1)}$ and $L_{(3, 1)}$, as shown in Table \ref{tableMp4}. As discussed in section \ref{claim}, $d_{(2,1)}$ distinguishes the two renormalization group flows inside $\mathcal{M}(p,4)$. Since the $\mathbb{Z}_2$ invertible symmetry is non-anomalous, $d_{(3,1)} =1$ for all $p$, and it is not useful as a constraint on the renormalization group flow.

\begin{table}[ht]
\begin{center}
\begin{tabular}{|c|c|c|c|c|c|c|c|c|}\hline
     $p$&3&5&7&9&11&13&15&17\\\hline
     $d_{(2, 1)}$&$\sqrt{2}$&$\sqrt{2}$&$- \sqrt{2}$&$- \sqrt{2}$&$\sqrt{2}$&$\sqrt{2}$&$- \sqrt{2}$&$- \sqrt{2}$\\\hline
     $d_{(3, 1)}$&1&1&1&1&1&1&1&1\\\hline
\end{tabular}
\caption{The quantum dimensions of $L_{(2, 1)} = \mathcal{N}$ and $L_{(3, 1)} = \eta$ in $\mathcal{M}(p , 4)$ series.}
\label{tableMp4}
\end{center}
\end{table}

We can also compute the spin contents of $L_{(2,1)}$ and $L_{(3,1)}$. Since the topological defect line $L_{(3,1)}$ generates the non-anomalous $\mathbb{Z}_2$ invertible symmetry, the spin contents must be $\{0, \pm \frac{1}{2} \}$ as can be checked easily, so we focus on $L_{(2,1)}$. The result of the explicit computation is summarized in Table \ref{table1}.

\begin{table}[ht]
\begin{center}
\scalebox{0.8}[0.8]{
\begin{tabular}{|c|c|c|}\hline
     $p$ & $h_{s, m} - h_{t, n}$ & the spin contents\\\hline
     3 & $\pm\frac{1}{16}, \pm\frac{7}{16}$ & $\pm\frac{1}{16}, \pm\frac{7}{16}$\\\hline
     5 & $\pm\frac{1}{16}, \pm\frac{7}{16}, \pm\frac{9}{16}, \pm\frac{17}{16}$ & $\pm\frac{1}{16}, \pm\frac{7}{16}$\\\hline
     7 & $\pm\frac{3}{16}, \pm\frac{5}{16}, \pm\frac{11}{16}, \pm\frac{13}{16}, \pm\frac{19}{16}, \pm\frac{27}{16}$ & $\pm\frac{3}{16}, \pm\frac{5}{16}$\\\hline
     9 & $\pm\frac{3}{16}, \pm\frac{5}{16}, \pm\frac{11}{16}, \pm\frac{13}{16}, \pm\frac{19}{16}, \pm\frac{21}{16}, \pm\frac{29}{16}, \pm\frac{37}{16}$ & $\pm\frac{3}{16}, \pm\frac{5}{16}$\\\hline
     11 & $\pm\frac{1}{16}, \pm\frac{7}{16}, \pm\frac{9}{16}, \pm\frac{15}{16}, \pm\frac{17}{16}, \pm\frac{23}{16}, \pm\frac{25}{16}, \pm\frac{31}{16}, \pm\frac{39}{16}, \pm\frac{47}{16}$ & $\pm\frac{1}{16}, \pm\frac{7}{16}$\\\hline
     13 & $\pm\frac{1}{16}, \pm\frac{7}{16}, \pm\frac{9}{16}, \pm\frac{15}{16}, \pm\frac{17}{16}, \pm\frac{23}{16}, \pm\frac{25}{16}, \pm\frac{31}{16}, \pm\frac{33}{16}, \pm\frac{41}{16}, \pm\frac{49}{16}, \pm\frac{57}{16}$ & $\pm\frac{1}{16}, \pm\frac{7}{16}$\\\hline
     15 & $\pm\frac{3}{16}, \pm\frac{5}{16}, \pm\frac{11}{16}, \pm\frac{13}{16}, \pm\frac{19}{16}, \pm\frac{21}{16}, \pm\frac{27}{16}, \pm\frac{29}{16}, \pm\frac{35}{16}, \pm\frac{37}{16}, \pm\frac{43}{16}, \pm\frac{51}{16}, \pm\frac{59}{16}, \pm\frac{67}{16}$ & $\pm\frac{3}{16}, \pm\frac{5}{16}$\\\hline
     17 & $\pm\frac{3}{16}, \pm\frac{5}{16}, \pm\frac{11}{16}, \pm\frac{13}{16}, \pm\frac{19}{16}, \pm\frac{21}{16}, \pm\frac{27}{16}, \pm\frac{29}{16}, \pm\frac{35}{16}, \pm\frac{37}{16}, \pm\frac{43}{16}, \pm\frac{45}{16}, \pm\frac{53}{16}, \pm\frac{61}{16}, \pm\frac{69}{16}, \pm\frac{77}{16}$ & $\pm\frac{3}{16}, \pm\frac{5}{16}$\\\hline
\end{tabular}}
\caption{The values of $h_{s, m} - h_{t, n}$ when ${N}_{(s, m)(t, n)}^
{(2, 1)} \neq 0$ and the spin content of the defect Hilbert space $\mathcal{H}_{\mathcal{N}}$ for minimal model $\mathcal{M}(p, 4)\ (p = 3, 5, \cdots, 17)$. The spin contents can be obtained by taking mod 1.} \label{table1}
\end{center}
\end{table}

As derived in section \ref{claim}, for $\mathcal{M}(p, 4)$, the general formula for $h_{s, m} - h_{t, n}$ and the spin content of $\mathcal{H}_{L_{(2,1)}}$ can be derived:
\begin{align}
    h_{s, m} - h_{t, n} : \Bigl\{\pm \dfrac{3p - 8m}{16}, \pm \dfrac{5p - 8m}{16}\ |\ 1 \le m \le p - 1\Bigr\},\\
    \text{the spin contents of}\ \mathcal{H}_{L_{(2,1)}} : \begin{cases}
        \pm\dfrac{1}{16}, \pm\dfrac{7}{16} & (p = 3, 5\ \text{mod}\ 8)\\
        \pm\dfrac{3}{16}, \pm\dfrac{5}{16} & (p = 1, 7\ \text{mod}\ 8)
    \end{cases}
\end{align}
This is compatible with (6.5) in \cite{Chang:2018iay} on the spin content of the duality defect line. With this agreement in mind, let us argue that $L_{(2,1)}$ is indeed the duality defect line.

As is well-known in the critical Ising model $\mathcal{M}(3,4)$ and the tricritical Ising model $\mathcal{M}(5,4)$, the $\mathbb{Z}_2$ Tambara-Yamagami modular tensor category has a physical interpretation as the duality defect line. We would like to offer a similar interpretation to the general $\mathcal{M}(p,4)$ Virasoro minimal models.

First, let us review the duality defect in more generality \cite{Choi:2021kmx, Choi:2022zal, Cui:2024cav}. Consider a quantum field theory $\mathcal{Q}$ on a $d$-dimensional spacetime with a non-anomalous $\mathbb{Z}_N$ $q$-form global symmetry. We divide the spacetime into two regions and gauge the $\mathbb{Z}_N$ global symmetry in half of the spacetime. Then, we can construct a codimension-one topological interface between the original theory $\mathcal{Q}$ and its gauged theory $\mathcal{Q}/\mathbb{Z}_N$. If the original theory $\mathcal{Q}$ is invariant under gauging the $\mathbb{Z}_N$ global symmetry
\begin{align}
    \mathcal{Q} \cong \mathcal{Q}/\mathbb{Z}_N \label{1}, 
\end{align}
the interface becomes the duality defect in the original theory $\mathcal{Q}$. 

When do we expect the duality defect in this so-called half-space gauging construction? Note that the gauged theory has $\widehat{\mathbb{Z}_N} \cong \mathbb{Z}_N$ $(d - q - 2)$-form dual (or quantum) symmetry, so the ungauged theory and the gauged theory may have the same symmetry  as in \eqref{1} if we require
\begin{align}
    q = \dfrac{d - 2}{2}.
\end{align}
Our cases discussed in this paper correspond to  $(d, q) = (2, 0)$.

Given a duality defect in the ultraviolet conformal field theory, we want to understand its fate under the renormalization group flow. Let us assume that the duality procedure commutes with the deformation of the ultraviolet conformal field theory.\footnote{This assumption is stronger than just assuming the deformation preserves $\mathbb{Z}_2$ invertible symmetry. For the difference, see the example of half-integer $k$ flows below.} Then, we should have the duality defect with the same property in the infrared as well. Such an existence should provide us with a strong constraint on the renormalization group flow.


We want to argue that the topological defect line $L_{(2,1)}$ in $\mathcal{M}(p,4)$ is the duality defect line. Let us consider general $\mathcal{M}(p,q)$ with the A-series modular invariant partition functions $Z_{A_{q - 1}, A_{p - 1}}$. Let us further assume the case when $p$ or $q$ is even (but not $2$) so that the $\mathbb{Z}_2$ symmetry is non-anomalous and can be gauged.
The partition function of the $\mathbb{Z}_2$ gauged theory can be obtained by the orbifold technique:
\begin{align}
    Z_{\mathcal{M}(p, q)/\mathbb{Z}_2} = \begin{cases}
        Z_{D_{q/2 + 1}, A_{p - 1}} & \text{when $q = 0$ mod 2}\\
        Z_{A_{q - 1}, D_{p/2 + 1}} & \text{when $p = 0$ mod 2}.
    \end{cases}
\end{align} 
The right-hand side is generically a non-diagonal modular invariant partition function of D-series, but the exception is when $q=4$ (or $p=4$). 
Because $A_3 \cong D_3$, A-series minimal models $\mathcal{M}(p, q)$ are invariant under gauging $\mathbb{Z}_2$ global symmetry if and only if $p$ or $q$ is 4. In other words,  $\mathcal{M}(4, q)$ is self-dual under the $\mathbb{Z}_2$ gauging and it has a duality defect line $\mathcal{N}$. Moreover, we can show that the duality defect line $\mathcal{N}$ satisfies the $\mathbb{Z}_2$ Tambara-Yamagami fusion rule 
 such as $\eta \times \eta = 1$, $\eta \times \mathcal{N} = \mathcal{N} \times \eta = \mathcal{N}$ and  $\mathcal{N} \times \mathcal{N} = 1+\eta $.
This is nothing but the fusion rule of the topological defect lines $L_{(2,1)}$ and $L_{(3,1)}$ preserved under the renormalization group flow $\mathcal{M}(4k+I,4) \to \mathcal{M}(4k-I,4)$ induced by $\phi_{(1,2k+1)}$.

After seeing how the duality line plays a significant role in the renormalization group flows of  $\mathcal{M}(4k+I,4) \to \mathcal{M}(4k-I,4)$ when $k$ is an integer, let us briefly discuss the case when $k$ is a half-integer. Here, $\phi_{(1,2k+1)}$ does not commute with $L_{(2,1)}$ and the only preserved symmetry is $L_{(3,1)}$. Accordingly, there are no obstructions for the renormalization group flows between any pair of $p$ in $\mathcal{M}(p,4)$. The simplest example would be $\mathcal{M}(3,4)$
 to ``$\mathcal{M}(1,4)$" induced by $\phi_{(1,2)} (= \epsilon)$, which is a massive flow. Physically, if we change the temperature of the Ising model, the Kramers–Wannier duality is broken and the criticality will be lost. See \cite{Dorey:2000zb} for an interpretation of ``$\mathcal{M}(1,4)$" from the viewpoint of the integral equations.

\begin{figure}[ht]
\centering
\begin{tikzpicture}
\draw(0, 0)node{$\mathcal{M}(3, 4)$};
\draw(2, 0)node{$\mathcal{M}(5, 4)$};
\draw(3, 2)node{$\mathcal{M}(7, 4)$};
\draw(5, 2)node{$\mathcal{M}(9, 4)$};
\draw(6, 0)node{$\mathcal{M}(11, 4)$};
\draw(8, 0)node{$\mathcal{M}(13, 4)$};
\draw(9, 2)node{$\mathcal{M}(15, 4)$};
\draw(11, 2)node{$\mathcal{M}(17, 4)$};
\draw[dashed, ->](3, 1.7)--(2, 0.3);
\draw(2.5, 1)node{\rotatebox{60}{$\times$}};
\draw[dashed, ->](6, 0.3)--(5, 1.7);
\draw(5.5, 1)node{\rotatebox{60}{$\times$}};
\draw[dashed, ->](9, 1.7)--(8, 0.3);
\draw(8.5, 1)node{\rotatebox{60}{$\times$}};
\draw[->](1.8, -0.3) to [out = 210, in = -30] (0.2, -0.3);
\draw(1, -1)node{$(k, I) = (1, 1)$};
\draw[->](7.8, -0.3) to [out = 210, in = -30] (6.2, -0.3);
\draw(7, -1)node{$(k, I) = (3, 1)$};
\draw[->](5.8, -0.3) to [out = 200, in = -20] (2.2, -0.3);
\draw(4, -1)node{$(k, I) = (2, 3)$};
\draw[->](4.8, 2.3) to [out = 150, in = 30] (3.2, 2.3);
\draw(4, 3)node{$(k, I) = (2, 1)$};
\draw[->](10.8, 2.3) to [out = 150, in = 30] (9.2, 2.3);
\draw(10, 3)node{$(k, I) = (4, 1)$};
\draw[->](8.8, 2.3) to [out = 160, in = 20] (5.2, 2.3);
\draw(7, 3)node{$(k, I) = (3, 3)$};
\draw(13, 1)node{$\cdots$};
\end{tikzpicture}
\caption{$\mathcal{M}(p, 4)$ are classified in terms of the spin contents and the quantum dimensions of the duality defect $\mathcal{N}$. They constrain the renormalization group flow like this figure. Dotted arrows are possible in the half-integer $k$ flow which does not preserve the duality defect lines.} 
\label{zu2}
\end{figure}
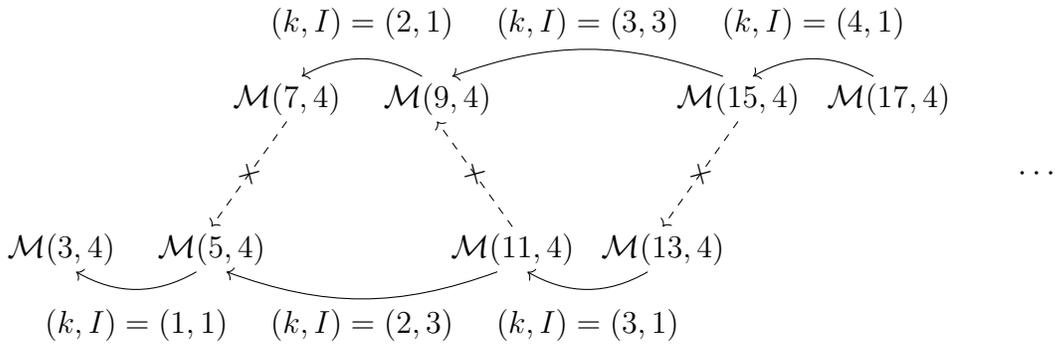

To conclude this subsection, we have summarized the renormalization group flows between $\mathcal{M}(p,4)$ in Figure \ref{zu2}. As a physical application beyond the classic tri-critical Ising to critical Ising flow $\mathcal{M}(5,4) \to \mathcal{M}(3,4)$, the fermionic version of the renormalization group flow $\mathcal{M}(11,4)$ to $\mathcal{M}(5,4)$ was discussed in \cite{Nakayama:2022svf} to understand the fate of the non-supersymmetric Yukawa fixed point.





\subsection{\texorpdfstring{$\mathcal{M}(5k+I,5) \to \mathcal{M}(5k-I,5)$}{TEXT} }

We repeat our analysis on the renormalization group flows $\mathcal{M}(5k+I,5) \to \mathcal{M}(5k-I,5)$ induced by $\phi_{(1,2k+1)}$ that preserve 
 a modular tensor category with the $SU(2)_3$ fusion ring, assuming $k$ is an integer. Here, we have $\varphi(5)=4$ distinct flows that cannot mix. The 't Hooft anomaly matching of the $\mathbb{Z}_2$ invertible symmetry distinguishes $\mathcal{M}(p,5) $ with even $p$ and odd $p$, but the other non-invertible symmetries give a finer classification.

The quantum dimensions of preserved non-invertible symmetries are given in Table \ref{tableMp5}. Here, $\phi = \frac{1+\sqrt{5}}{2}$ is the golden ratio. As discussed in section \ref{claim}, not only all the quantum dimensions are consistent with the proposed renormalization group flow, but $L_{(2,1)}$ gives the strongest constraint such that it fully distinguishes our proposed renormalization group flows.
\begin{table}[ht]
\begin{center}
\begin{tabular}{|c|c|c|c|c|c|c|c|c|c|}\hline
     $p$&2&3&4&6&7&8&9&11& $\cdots$ \\\hline
$d_{(2, 1)}$&$-\phi^{-1} $&$\phi^{-1} $&$\phi$&$ \phi $&$\phi^{-1}$&$-\phi^{-1} $&$-\phi $&$-\phi$ & $\cdots $\\\hline 
$d_{(3, 1)}$&$-\phi^{-1} $&$-\phi^{-1} $&$\phi$&$ \phi $&$-\phi^{-1}$&$-\phi^{-1} $&$\phi $&$\phi$ & $\cdots $\\\hline 
     $d_{(4, 1)}$&$1$&$-1$&$1$&$1$&$-1$&$1$&$-1$&$-1$ & $\cdots$ \\\hline
\end{tabular}
\caption{The quantum dimensions of $L_{(r,1)} $ in $\mathcal{M}(p , 5)$ series.}\label{tableMp5}
\end{center}
\end{table}

Let us also compute the spin contents of the preserved topological defect lines. On  $\mathcal{H}_{L_{(2,1)}}$, the non-zero fusion coefficients are $N_{(1, m)(2, m)}^{(2, 1)}$, $N_{(2, m)(1, m)}^{(2, 1)}$,  $N_{(2, m)(3, m)}^{(2, 1)}$, $N_{(3, m)(2, m)}^{(2, 1)}$, $N_{(3, m)(4, m)}^{(2, 1)}$, and $N_{(4, m)(3, m)}^{(2, 1)}$, so we have
\begin{align}
\left\{ \pm \left(\frac{3p}{20} -\frac{m}{2}\right), \pm \frac{1}{4}(p-2m), \pm \left(\frac{7p}{20} -\frac{m}{2}\right)  \right\} ,
\end{align}
where $1 \le m \le p - 1$. Similarly on $\mathcal{H}_{L_{(3,1)}}$, from the non-zero fusion coefficients $N_{(1, m)(3, m)}^{(3, 1)}$, $N_{(3, m)(1, m)}^{(3, 1)}$, $N_{(2, m)(4, m)}^{(3, 1)}$, $N_{(4, m)(2, m)}^{(3, 1)}$, $N_{(2, m)(2, m)}^{(3, 1)}$, and $N_{(3, m)(3, m)}^{(3, 1)}$, we have
\begin{align}
\left\{ \pm \left(\frac{2p}{5}-m\right), \pm \left(\frac{3p}{5} -m\right), 0 \right\} .
\end{align}
Finally, on $\mathcal{H}_{L_{(4,1)}}$, from the non-zero fusion coefficients $N_{(1, m)(4, m)}^{(4, 1)}$, $N_{(4, m)(1, m)}^{(4, 1)}$, $N_{(2, m)(3, m)}^{(4, 1)}$, and $N_{(3, m)(2, m)}^{(4, 1)}$,  we have 
\begin{align}
\left\{ \pm \frac{3}{4}(p-2m), \pm \frac{1}{4}(p-2m) \right\} .
\end{align}

The results are summarized in Table \ref{table4}. We note that the constraints from the quantum dimensions and constraints from the spin contents are the same.

\begin{table}[ht]
\begin{center}
\scalebox{0.65}[0.75]{
\begin{tabular}{|c|c|c|c|c|c|c|c|c|}\hline
     $p$&2&3&4&6&7&8&9&11\\\hline
     $\mathcal{H}_{L_{(2, 1)}}$&$0, \pm\frac{1}{5}$&$\pm\frac{1}{20}, \pm\frac{1}{4}, \pm\frac{9}{20}$&$0, \pm\frac{1}{10}, \pm\frac{2}{5}, \pm\frac{1}{2}$&$0, \pm\frac{1}{10}, \pm\frac{2}{5}, \pm\frac{1}{2}$&$\pm\frac{1}{20}, \pm\frac{1}{4}, \pm\frac{9}{20}$&$0, \pm\frac{1}{5}, \pm\frac{3}{10}, \pm\frac{1}{2}$&$\pm\frac{3}{20}, \pm\frac{1}{4}, \pm\frac{7}{20}$&$\pm\frac{3}{20}, \pm\frac{1}{4}, \pm\frac{7}{20}$\\\hline
     $\mathcal{H}_{L_{(3, 1)}}$&$0, \pm\frac{1}{5}$&$0, \pm\frac{1}{5}$&$0, \pm\frac{2}{5}$&$0, \pm\frac{2}{5}$&$0, \pm\frac{1}{5}$&$0, \pm\frac{1}{5}$&$0, \pm\frac{2}{5}$&$0, \pm\frac{2}{5}$\\\hline
     $\mathcal{H}_{L_{(4, 1)}}$&$0$&$\pm\frac{1}{4}$&$0, \pm\frac{1}{2}$&$0, \pm\frac{1}{2}$&$\pm\frac{1}{4}$&$0, \pm\frac{1}{2}$&$\pm\frac{1}{4}$&$\pm\frac{1}{4}$\\\hline
\end{tabular}}
\caption{The spin contents of $\mathcal{H}_{L_{(r, 1)}}$ in $\mathcal{M}(p , 5)$ series.}\label{table4}
\end{center}
\end{table}

Since this is the first example of the appearance of $p=2$ as in the flow $\mathcal{M}(8,5) \to \mathcal{M}(2,5)$ induced by $\phi_{(1, 3)}$, let us comment on it. We will give more detailed discussions in section \ref{M2}. Unlike in $\mathcal{M}(p,5)$ with $p>2$, in $\mathcal{M}(2,5)$, $L_{(4,1)}$ is identified with $L_{(1,1)}$ and $L_{(3,1)}$ is identified with $L_{(2,1)}$, so we have only half of the preserved non-invertible symmetries in the infrared compared with the ones in the ultraviolet. These symmetries may be spontaneously broken or disappear. Note, in particular, that The $\mathbb{Z}_2$ invertible symmetry generated by $L_{(4,1)}$ is non-anomalous, so it can simply disappear. Understanding the fate of these symmetries requires dynamical analysis which is beyond the scope of our paper, but further discussions follow in section \ref{M2}.

Let us finally make a brief discussion on the half-integer $k$ flow. The preserved topological defect lines are $L_{(1,1)}$ and $L_{(3,1)}$, which form the fusion rule of $\mathcal{M}(2,5)$ (Lee-Yang fusion ring). It is also called the Fibonacci fusion ring (see the appearance of the golden ratio above). There exist two distinct renormalization group flows depending on $d_{(3,1)}$, which is consistent with the classification of the modular tensor category with the Fibonacci fusion ring.
The $\mathbb{Z}_2$ invertible symmetry is broken by the deformation, but it reappears as an emergent symmetry. This is how we can change the (non-)anomalous nature of the $\mathbb{Z}_2$ invertible symmetry along the renormalization group flow.

\subsection{Flows to \texorpdfstring{$\mathcal{M}(p,2)$ or $\mathcal{M}(2,q)$ }{TEXT}}\label{M2}
As far as the non-invertible symmetry goes, there is nothing wrong with our flows $\mathcal{M}(kq + I, q) \to\mathcal{M}(kq-I, q)$ even when $q=2$ or $kq-I = 2$. A closer look tells us, however, the situation is more subtle essentially because there is no $\mathbb{Z}_2$ invertible symmetry in $\mathcal{M}(p,q)$ when either $p$ or $q$ is $2$. From the viewpoint of the topological defect lines, the would-be  $\mathbb{Z}_2$ topological defect line $L_{(q - 1, 1)}$ is identified with the identity defect line $L_{(1, 1)}$. In other words, there exist no charged primary operators under the $\mathbb{Z}_2$ invertible symmetry.

By taking $q = 2$, our proposal suggests the existence of the renormalization group flow $\mathcal{M}(2k + I, 2) \to \mathcal{M}(2k - I, 2)$ induced by $\phi_{(1, 2k + 1)}$. The case with $I=1$, however, may sound problematic because  $\phi_{(1,2k + 1)}$ is outside of the Kac table. The most reasonable interpretation is that the flow exists but the primary operator that induces the renormalization group flow is a certain fine-tuned combination of $\phi_{(1, l)}$ with $l=1,2, \cdots, k$.\footnote{Under the identification, this is equivalent to taking $l=1,3, \cdots, 2k-1$.} Indeed, the numerical analysis based on the truncated conformal space approach shows the existence of such flows (for small $k$) \cite{Lencses:2022ira,Lencses:2023evr,Lencses:2024wib}. Of course, we could be simply agnostic about these flows because the argument based on the non-invertible symmetries does not give any constraint at all here, but we are tempted to unify them within our proposal.

Another subtle case is $q\neq 2$ but $kq - I = 2$. In this case, the question is about the fate of the $\mathbb{Z}_2$ invertible symmetry.
We start with the ultraviolet theory with a $\mathbb{Z}_2$ invertible symmetry and we add the deformation that preserves it. In the infrared limit, we end up with the theory that does {\it not} possess any $\mathbb{Z}_2$ invertible symmetry. What happened to the symmetry? One possibility is it is spontaneously broken. Another possibility is all the charged operators are decoupled in the infrared limit. 
 The latter is possible because the $\mathbb{Z}_2$ symmetry here is non-anomalous.
 
 In the massive flow example of $\mathcal{M}(4,3) \to \mathcal{M}(2,3)$ induced by $\phi_{(1,3)}$ (formally $k=1$, $I=1$ flow), where a Lagrangian description is available, both can happen depending on the sign of the deformation. Regarding $\mathcal{M}(4,3) $ as a critical $\phi^4$ theory, adding $m^2\phi^2$ with positive $m^2$ gives decoupling of the $\mathbb{Z}_2$ charged field $\phi$, and with negative $m^2$ it gives spontaneous breaking of the $\mathbb{Z}_2$ symmetry.

Physically $\mathcal{M}(2, q)$ can be realized in the multi-critical Lee-Yang fixed point in two dimensions.
So far, the canonical construction of the (multi-critical) Lee-Yang fixed point was to start with the $\mathbb{Z}_2$ symmetric theory, say Ising model, and add the (imaginary) $\mathbb{Z}_2$ breaking interaction, say pure imaginary magnetic field, to get to the fixed point. This new construction without explicitly breaking the $\mathbb{Z}_2$ symmetry is novel and worth studying further.

\section{Discussions}\label{Conclusion}

In this paper, based on the study of non-invertible symmetries, we have proposed there exist infinitely many new renormalization group flows between Virasoro minimal models  $\mathcal{M}(kq + I, q) \to\mathcal{M}(kq-I, q)$ induced by $\phi_{(1,2k+1)}$. They vastly generalize the previously proposed renormalization group flows in the literature.

One should be able to check our new renormalization group flows by using the conformal perturbation theory or truncated conformal space approach \cite{Yurov:1989yu,Yurov:1991my,Feverati:2006ni}. Here, let us simply observe that the deformation operator $\phi_{(1,2k+1)}$ is more relevant when the separation of the renormalization group flow given by  $I$ is larger. This agrees with our intuitive picture; the more relevant deformations, the longer the renormalization group trajectory becomes. It should also agree with the difference of the central charge at the fixed points. More recent approaches to the renormalization group flow based on the boundary states and interfaces can be found in \cite{Gaiotto:2012np,Cardy:2017ufe,Konechny:2023xvo}, and they may be useful to understand our new renormalization group flows better (see also the recent work \cite{Cogburn:2023xzw}).

For a future direction, it is interesting to study the integrable structure of our new renormalization group flows  $\mathcal{M}(kq + I, q) \to\mathcal{M}(kq-I, q)$. As initiated in \cite{Zamolodchikov:1987jf,Zamolodchikov:1991vx} the flows with $I=1$ (or $I=\frac{1}{2})$ can be described by the Thermodynamic Bethe Ansatz \cite{Zamolodchikov:1989cf,Zamolodchikov:1991vh,Bazhanov:1994ft,Feverati:1995hy,Kausch:1996vq,Pearce:2000dv,Pearce:2003km}. When $I>1$, there seem to be no known descriptions in terms of the Thermodynamic Bethe Ansatz, but the integral equations that describe the effective central charge along the flow were proposed 
 \cite{Dorey:2000zb}. It should be interesting to generalize these works to our new renormalization group flows.

Another direction to be explored further is a concrete realization of our new renormalization group flows in statistical models or quantum field theories. It is known that any (non-unitary) minimal models have a statistical model realization based on the Restricted Solid-On-Solid (RSOS) model \cite{Pasquier:1987tm,Nakanishi:1989cv,Takacs:1996wt,Takacs:2002fg} (see also \cite{Bianchini:2015zka,2017PhRvL.119d0601C} for the related spin chain construction). It is interesting to understand the nature of the anomalous $\mathbb{Z}_2$ symmetries, duality defects as well as general non-invertible symmetries preserved by our renormalization group flow.

As for the quantum field theory realization, we are still in our infancy even more than fifty years after the birth of conformal field theories. This is because the quantum field theory descriptions (e.g. Landau-Ginzburg theory \cite{Zamolodchikov:1986db} or gauge theory \cite{Delmastro:2021otj}) for non-unitary minimal models are largely unknown. We, nevertheless, note that $2d$-$4d$ correspondence \cite{Beem:2017ooy} gives realizations of certain (but probably not all) non-unitary chiral algebra from supersymmetric conformal field theories in four dimensions. 
We hope our renormalization group flow may become a breakthrough in this direction. 

Concerning the explicit constructions of minimal models from known quantum field theories, while our discussions focus on the infrared flow, it seems interesting to look at the flow in the opposite way as an ultraviolet flow \cite{Ahn:2022pia,Ahn:2024sxt}. Since the ultraviolet flow is a non-renormalizable deformation, we may need some other input such as an integrability to pursue, but once done, it should give a strong hint toward the quantum field theory descriptions of the ultraviolet theory whose ``Lagrangian description" were unknown.

Why do we crave Lagrangian descriptions of the Virasoro minimal models?
We wish to point out here that once the Lagrangian descriptions are available, we may generalize these models in higher dimensions. When we obtain the higher dimensional fixed points, we can ask whether the constraints on the renormalization flow from the non-invertible symmetries can also be uplifted. Since we have less explicit constructions of non-invertible symmetries in higher dimensions, this is an outstanding question to be pursued.

\section*{Acknowledgments}
TT is supported by JSPS KAKENHI Grant Number JP24KJ1500.
YN is in part supported by JSPS KAKENHI Grant Number 21K03581. YN would like to thank C.~Ahn and K.~Kikuchi for the correspondence and discussions.

\bibliography{minimalmodelRGflow}{}

\begin{thebibliography}{10}

\bibitem{Zamolodchikov:1987ti}
A.~B. Zamolodchikov,
\newblock Sov. J. Nucl. Phys. {\bf 46}, 1090 (1987).

\bibitem{Ahn:1992qi}
C.-r. Ahn,
\newblock Phys. Lett. B {\bf 294}, 204 (1992), hep-th/9202028.

\bibitem{Lassig:1991an}
M.~Lassig,
\newblock Phys. Lett. B {\bf 278}, 439 (1992).

\bibitem{Martins:1992ht}
M.~J. Martins,
\newblock Phys. Rev. Lett. {\bf 69}, 2461 (1992), hep-th/9205024.

\bibitem{Martins:1992yk}
M.~J. Martins,
\newblock Nucl. Phys. B {\bf 394}, 339 (1993), hep-th/9208011.

\bibitem{Dorey:2000zb}
P.~Dorey, C.~Dunning, and R.~Tateo,
\newblock Nucl. Phys. B {\bf 578}, 699 (2000), hep-th/0001185.

\bibitem{Klebanov:2022syt}
I.~R. Klebanov, V.~Narovlansky, Z.~Sun, and G.~Tarnopolsky,
\newblock JHEP {\bf 02}, 066 (2023), 2211.07029.

\bibitem{Lencses:2022ira}
M.~Lencs\'es, A.~Miscioscia, G.~Mussardo, and G.~Tak\'acs,
\newblock JHEP {\bf 02}, 046 (2023), 2211.01123.

\bibitem{Lencses:2023evr}
M.~Lencs\'es, A.~Miscioscia, G.~Mussardo, and G.~Tak\'acs,
\newblock JHEP {\bf 09}, 052 (2023), 2304.08522.

\bibitem{Chang:2018iay}
C.-M. Chang, Y.-H. Lin, S.-H. Shao, Y.~Wang, and X.~Yin,
\newblock JHEP {\bf 01}, 026 (2019), 1802.04445.

\bibitem{Lin:2019hks}
Y.-H. Lin and S.-H. Shao,
\newblock J. Phys. A {\bf 54}, 065201 (2021), 1911.00042.

\bibitem{Aasen:2020jwb}
D.~Aasen, P.~Fendley, and R.~S.~K. Mong,
\newblock (2020), 2008.08598.

\bibitem{Thorngren:2021yso}
R.~Thorngren and Y.~Wang,
\newblock JHEP {\bf 07}, 051 (2024), 2106.12577.

\bibitem{Huang:2021nvb}
T.-C. Huang, Y.-H. Lin, K.~Ohmori, Y.~Tachikawa, and M.~Tezuka,
\newblock Phys. Rev. Lett. {\bf 128}, 231603 (2022), 2110.03008.

\bibitem{Buican:2021uyp}
M.~Buican, A.~Dymarsky, and R.~Radhakrishnan,
\newblock JHEP {\bf 03}, 017 (2023), 2112.12162.

\bibitem{Burbano:2021loy}
I.~M. Burbano, J.~Kulp, and J.~Neuser,
\newblock JHEP {\bf 10}, 186 (2022), 2112.14323.

\bibitem{Chang:2022hud}
C.-M. Chang, J.~Chen, and F.~Xu,
\newblock SciPost Phys. {\bf 15}, 216 (2023), 2208.02757.

\bibitem{Lin:2022dhv}
Y.-H. Lin, M.~Okada, S.~Seifnashri, and Y.~Tachikawa,
\newblock JHEP {\bf 03}, 094 (2023), 2208.05495.

\bibitem{Lu:2022ver}
D.-C. Lu and Z.~Sun,
\newblock JHEP {\bf 02}, 173 (2023), 2208.06077.

\bibitem{Kaidi:2022cpf}
J.~Kaidi, K.~Ohmori, and Y.~Zheng,
\newblock Commun. Math. Phys. {\bf 404}, 1021 (2023), 2209.11062.

\bibitem{Cheng:2022sgb}
M.~Cheng and N.~Seiberg,
\newblock SciPost Phys. {\bf 15}, 051 (2023), 2211.12543.

\bibitem{Chatterjee:2022jll}
A.~Chatterjee, W.~Ji, and X.-G. Wen,
\newblock (2022), 2212.14432.

\bibitem{Lin:2023uvm}
Y.-H. Lin and S.-H. Shao,
\newblock Phys. Rev. D {\bf 107}, 125025 (2023), 2302.13900.

\bibitem{Jacobsen:2023isq}
J.~L. Jacobsen and H.~Saleur,
\newblock JHEP {\bf 12}, 090 (2023), 2305.05746.

\bibitem{Choi:2023xjw}
Y.~Choi, B.~C. Rayhaun, Y.~Sanghavi, and S.-H. Shao,
\newblock Phys. Rev. D {\bf 108}, 125005 (2023), 2305.09713.

\bibitem{Haghighat:2023sax}
B.~Haghighat and Y.~Sun,
\newblock (2023), 2306.16555.

\bibitem{Seiberg:2023cdc}
N.~Seiberg and S.-H. Shao,
\newblock SciPost Phys. {\bf 16}, 064 (2024), 2307.02534.

\bibitem{Antinucci:2023ezl}
A.~Antinucci, F.~Benini, C.~Copetti, G.~Galati, and G.~Rizi,
\newblock (2023), 2308.11707.

\bibitem{Duan:2023ykn}
Z.~Duan, Q.~Jia, and S.~Lee,
\newblock JHEP {\bf 11}, 206 (2023), 2309.01913.

\bibitem{Chen:2023jht}
J.~Chen, B.~Haghighat, and Q.-R. Wang,
\newblock (2023), 2309.01914.

\bibitem{Nagoya:2023zky}
Y.~Nagoya and S.~Shimamori,
\newblock JHEP {\bf 12}, 062 (2023), 2309.05294.

\bibitem{Sinha:2023hum}
M.~Sinha, F.~Yan, L.~Grans-Samuelsson, A.~Roy, and H.~Saleur,
\newblock (2023), 2310.19703.

\bibitem{Choi:2023vgk}
Y.~Choi, D.-C. Lu, and Z.~Sun,
\newblock JHEP {\bf 01}, 142 (2024), 2310.19867.

\bibitem{Diatlyk:2023fwf}
O.~Diatlyk, C.~Luo, Y.~Wang, and Q.~Weller,
\newblock JHEP {\bf 03}, 127 (2024), 2311.17044.

\bibitem{Cordova:2023qei}
C.~Cordova and G.~Rizi,
\newblock (2023), 2312.17308.

\bibitem{Grover:2023loq}
S.~Grover, S.~Hegde, and D.~P. Jatkar,
\newblock JHEP {\bf 05}, 057 (2024), 2312.17165.

\bibitem{Seiberg:2024gek}
N.~Seiberg, S.~Seifnashri, and S.-H. Shao,
\newblock SciPost Phys. {\bf 16}, 154 (2024), 2401.12281.

\bibitem{Copetti:2024rqj}
C.~Copetti, L.~Cordova, and S.~Komatsu,
\newblock (2024), 2403.04835.

\bibitem{Chatterjee:2024ych}
A.~Chatterjee, O.~M. Aksoy, and X.-G. Wen,
\newblock (2024), 2405.05331.

\bibitem{Nakayama:2022svf}
Y.~Nakayama and K.~Kikuchi,
\newblock JHEP {\bf 03}, 240 (2023), 2212.06342.

\bibitem{DiFrancesco:1997nk}
P.~Di~Francesco, P.~Mathieu, and D.~Senechal,
\newblock {\em {Conformal Field Theory}}Graduate Texts in Contemporary Physics (Springer-Verlag, New York, 1997).

\bibitem{Petkova:1988cy}
V.~B. Petkova,
\newblock Int. J. Mod. Phys. A {\bf 3}, 2945 (1988).

\bibitem{Hsieh:2020uwb}
C.-T. Hsieh, Y.~Nakayama, and Y.~Tachikawa,
\newblock Phys. Rev. Lett. {\bf 126}, 195701 (2021), 2002.12283.

\bibitem{Kikuchi:2021qxz}
K.~Kikuchi,
\newblock (2021), 2109.02672.

\bibitem{Kikuchi:2022jbl}
K.~Kikuchi,
\newblock (2022), 2204.03247.

\bibitem{Kikuchi:2022gfi}
K.~Kikuchi,
\newblock (2022), 2207.06433.

\bibitem{Kikuchi:2022biw}
K.~Kikuchi,
\newblock (2022), 2207.10095.

\bibitem{Kikuchi:2022rco}
K.~Kikuchi,
\newblock (2022), 2209.00016.

\bibitem{Castro-Alvaredo:2017udm}
O.~A. Castro-Alvaredo, B.~Doyon, and F.~Ravanini,
\newblock J. Phys. A {\bf 50}, 424002 (2017), 1706.01871.

\bibitem{Bender:2023cem}
C.~M. Bender and D.~W. Hook,
\newblock (2023), 2312.17386.

\bibitem{Gukov:2015qea}
S.~Gukov,
\newblock JHEP {\bf 01}, 020 (2016), 1503.01474.

\bibitem{Gukov:2016tnp}
S.~Gukov,
\newblock Nucl. Phys. B {\bf 919}, 583 (2017), 1608.06638.

\bibitem{Zamolodchikov:1987jf}
A.~B. Zamolodchikov,
\newblock JETP Lett. {\bf 46}, 160 (1987).

\bibitem{Ravanini:1994pt}
F.~Ravanini, M.~Stanishkov, and R.~Tateo,
\newblock Int. J. Mod. Phys. A {\bf 11}, 677 (1996), hep-th/9411085.

\bibitem{Fei:2014xta}
L.~Fei, S.~Giombi, I.~R. Klebanov, and G.~Tarnopolsky,
\newblock Phys. Rev. D {\bf 91}, 045011 (2015), 1411.1099.

\bibitem{Delouche:2023wsl}
O.~Delouche, J.~Elias~Miro, and J.~Ingoldby,
\newblock SciPost Phys. {\bf 16}, 105 (2024), 2312.09221.

\bibitem{Choi:2021kmx}
Y.~Choi, C.~Cordova, P.-S. Hsin, H.~T. Lam, and S.-H. Shao,
\newblock Phys. Rev. D {\bf 105}, 125016 (2022), 2111.01139.

\bibitem{Choi:2022zal}
Y.~Choi, C.~Cordova, P.-S. Hsin, H.~T. Lam, and S.-H. Shao,
\newblock Commun. Math. Phys. {\bf 402}, 489 (2023), 2204.09025.

\bibitem{Cui:2024cav}
W.~Cui, B.~Haghighat, and L.~Ruggeri,
\newblock (2024), 2406.09261.

\bibitem{Lencses:2024wib}
M.~Lencs\'es, A.~Miscioscia, G.~Mussardo, and G.~Tak\'acs,
\newblock (2024), 2404.06100.

\bibitem{Yurov:1989yu}
V.~P. Yurov and A.~B. Zamolodchikov,
\newblock Int. J. Mod. Phys. A {\bf 5}, 3221 (1990).

\bibitem{Yurov:1991my}
V.~P. Yurov and A.~B. Zamolodchikov,
\newblock Int. J. Mod. Phys. A {\bf 6}, 4557 (1991).

\bibitem{Feverati:2006ni}
G.~Feverati, K.~Graham, P.~A. Pearce, G.~Z. Toth, and G.~Watts,
\newblock J. Stat. Mech. {\bf 0803}, P03011 (2008), hep-th/0612203.

\bibitem{Gaiotto:2012np}
D.~Gaiotto,
\newblock JHEP {\bf 12}, 103 (2012), 1201.0767.

\bibitem{Cardy:2017ufe}
J.~Cardy,
\newblock SciPost Phys. {\bf 3}, 011 (2017), 1706.01568.

\bibitem{Konechny:2023xvo}
A.~Konechny,
\newblock JHEP {\bf 11}, 004 (2023), 2306.13719.

\bibitem{Cogburn:2023xzw}
C.~V. Cogburn, A.~L. Fitzpatrick, and H.~Geng,
\newblock SciPost Phys. Core {\bf 7}, 021 (2024), 2308.00737.

\bibitem{Zamolodchikov:1991vx}
A.~B. Zamolodchikov,
\newblock Nucl. Phys. B {\bf 358}, 524 (1991).

\bibitem{Zamolodchikov:1989cf}
A.~B. Zamolodchikov,
\newblock Nucl. Phys. B {\bf 342}, 695 (1990).

\bibitem{Zamolodchikov:1991vh}
A.~B. Zamolodchikov,
\newblock Nucl. Phys. B {\bf 358}, 497 (1991).

\bibitem{Bazhanov:1994ft}
V.~V. Bazhanov, S.~L. Lukyanov, and A.~B. Zamolodchikov,
\newblock Commun. Math. Phys. {\bf 177}, 381 (1996), hep-th/9412229.

\bibitem{Feverati:1995hy}
G.~Feverati, E.~Quattrini, and F.~Ravanini,
\newblock Phys. Lett. B {\bf 374}, 64 (1996), hep-th/9512104.

\bibitem{Kausch:1996vq}
H.~Kausch, G.~Takacs, and G.~Watts,
\newblock Nucl. Phys. B {\bf 489}, 557 (1997), hep-th/9605104.

\bibitem{Pearce:2000dv}
P.~A. Pearce, L.~Chim, and C.-r. Ahn,
\newblock Nucl. Phys. B {\bf 601}, 539 (2001), hep-th/0012223.

\bibitem{Pearce:2003km}
P.~A. Pearce, L.~Chim, and C.~Ahn,
\newblock Nucl. Phys. B {\bf 660}, 579 (2003), hep-th/0302093.

\bibitem{Pasquier:1987tm}
V.~Pasquier,
\newblock Nucl. Phys. B {\bf 295}, 491 (1988).

\bibitem{Nakanishi:1989cv}
T.~Nakanishi,
\newblock Nucl. Phys. B {\bf 334}, 745 (1990).

\bibitem{Takacs:1996wt}
G.~Takacs,
\newblock Nucl. Phys. B {\bf 489}, 532 (1997), hep-th/9604098.

\bibitem{Takacs:2002fg}
G.~Takacs and G.~Watts,
\newblock Nucl. Phys. B {\bf 642}, 456 (2002), hep-th/0203073.

\bibitem{Bianchini:2015zka}
D.~Bianchini and F.~Ravanini,
\newblock Jounral of Physics A  (2015), 1509.04601.

\bibitem{2017PhRvL.119d0601C}
R.~{Couvreur}, J.~L. {Jacobsen}, and H.~{Saleur},
\newblock Phys. Rev. Lett. {\bf 119}, 040601 (2017), 1611.08506.

\bibitem{Zamolodchikov:1986db}
A.~B. Zamolodchikov,
\newblock Sov. J. Nucl. Phys. {\bf 44}, 529 (1986).

\bibitem{Delmastro:2021otj}
D.~Delmastro, J.~Gomis, and M.~Yu,
\newblock JHEP {\bf 02}, 157 (2023), 2108.02202.

\bibitem{Beem:2017ooy}
C.~Beem and L.~Rastelli,
\newblock JHEP {\bf 08}, 114 (2018), 1707.07679.

\bibitem{Ahn:2022pia}
C.~Ahn and A.~LeClair,
\newblock JHEP {\bf 08}, 179 (2022), 2205.10905.

\bibitem{Ahn:2024sxt}
C.~Ahn and Z.~Bajnok,
\newblock (2024), 2407.06582.

\end{thebibliography}
\bibliographystyle{h-physrev}

\end{document}